\documentclass[conference]{IEEEtran}
\IEEEoverridecommandlockouts

\usepackage[acronym]{glossaries}
\newacronym{BS}{BS}{base station}
\newacronym{PS}{PS}{phase-shifter}
\newacronym{RL}{RL}{reinforcement learning}
\newacronym{AP}{AP}{analog precoder}
\newacronym{FC-HBF}{FC-HBF}{fully-connected HBF}
\newacronym{FSA-HBF}{FSA-HBF}{fixed subarray HBF}
\newacronym{DSA-HBF}{DSA-HBF}{dynamic subarray HBF}
\newacronym{BF}{BF}{beamforming}
\newacronym{UE}{UE}{user equipment}
\newacronym{AWGN}{AWGN}{additive white gaussian noise}
\newacronym{MIMO}{MIMO}{multiple-input multiple-output}
\newacronym{MISO}{MISO}{multiple-input single-output}
\newacronym{RF}{RF}{radio frequency}
\newacronym{RIS}{RIS}{reconfigurable intelligent surfaces}
\newacronym{IOT}{IOT}{internet-of-things}
\newacronym{CL}{CL}{convolutional layer}
\newacronym{FDD}{FDD}{frequency division duplex}
\newacronym{TDD}{TDD}{time division duplex}
\newacronym{CSI}{CSI}{channel state information}
\newacronym{DNN}{DNN}{deep neural network}
\newacronym{DP}{DP}{digital precoder}
\newacronym{DL}{DL}{deep learning}
\newacronym{SVD}{SVD}{singular-value decomposition}
\newacronym{CNN}{CNN}{convolution neural network}
\newacronym{FDP}{FDP}{fully digital precoder}
\newacronym{SE}{SE}{spectral efficiency}
\newacronym{OFDM}{OFDM}{orthogonal frequency division multiplexing}
\newacronym{OMP}{OMP}{orthogonal matching pursuit}
\newacronym{FL}{FL}{fully-connected layer}
\newacronym{HSHO}{HSHO}{Hybrid Structured Heuristic Optimization}
\newacronym{HBF}{HBF}{hybrid beamforming}
\newacronym{IA}{IA}{initial access}
\newacronym{mm-Wave}{mm-Wave}{millimeter wave}
\newacronym{mMIMO}{mMIMO}{massive MIMO}
\newacronym{SINR}{SINR}{signal-to-interference-noise ratio}
\newacronym{SNR}{SNR}{signal-to-noise ratio}
\newacronym{RSSI}{RSSI}{received signal strength indicator}
\newacronym{PZF}{PZF}{phase zero forcing}
\newacronym{PSO}{PSO}{particle swarm optimization}
\newacronym{ZF}{ZF}{zero forcing}
\newacronym{O-FDP}{O-FDP}{optimal fully digital precoder}
\newacronym{JT}{JT}{joint transmission}
\newacronym{CU}{CU}{central unit}
\newacronym{MSE}{MSE}{mean square error}
\newacronym{CEL}{CEL}{cross entropy loss}
\newacronym{CB}{CB}{conjugate beamforming}
\newacronym{NC}{NC}{network controller}
\newacronym{CoMP}{CoMP}{coordinated multi point}
\newacronym{CF-mMIMO}{CF-mMIMO}{cell-free massive MIMO}
\newacronym{CF-HBF}{CF-HBF}{cell-free hybrid beamforming}
\newacronym{CF-BF}{CF-BF}{cell-free beamforming}

\newif\ifDeepMIMOModel
\DeepMIMOModeltrue

\newif\ifSimpleNParamEq
\SimpleNParamEqtrue

\usepackage{lettrine}

\usepackage{ifpdf}

\ifCLASSINFOpdf
  \usepackage[pdftex]{graphicx}
  \usepackage{graphicx,epstopdf}
  \graphicspath{{../pdf/}{../jpeg/}}
  \DeclareGraphicsExtensions{.pdf,.jpeg,.png,.eps}
\else
  \usepackage[dvips]{graphicx}
  \usepackage{graphicx,epstopdf}
  \graphicspath{{../eps/}}
  \DeclareGraphicsExtensions{}
\fi

\usepackage{cite}
\usepackage{algorithmic}
\usepackage{amsthm}
\usepackage{steinmetz}
\usepackage{amssymb}
\usepackage{balance}
\usepackage{eqparbox}
\usepackage{multirow}
\usepackage{bbm}

\usepackage[ruled,vlined]{algorithm2e}
\SetAlFnt{\small}

\usepackage{amsmath}

\usepackage{mathtools, nccmath}

\usepackage{booktabs, siunitx}
\usepackage[svgnames,table]{xcolor}
\usepackage[scaled]{DejaVuSansMono}
\usepackage[T1]{fontenc}

\usepackage{color}
\usepackage{relsize}
\usepackage{mathtools}

\newcommand{\bs}[1]{\boldsymbol{#1}}

\newcommand{\mb}[1]{\mathbf{#1}}
\newcommand{\mr}[1]{\mathrm{#1}}

\usepackage{mathtools}

\DeclareMathOperator*{\argmax}{arg\;max}

\DeclareMathOperator*{\minimize}{minimize}

\SetKwInput{KwInput}{Input}                
\SetKwInput{KwOutput}{Output}
\SetKwInput{KwOutputr}{Output Regression}              
\SetKwInput{KwOutputc}{Output Classification}              

\newcommand{\bseq}{\begin{subequations}}
\newcommand{\eseq}{\end{subequations}}
\newcommand{\baln}{\begin{align}}
\newcommand{\ealn}{\end{align}}
\newcommand{\balnd}{\begin{aligned}}
\newcommand{\ealnd}{\end{aligned}}
\newcommand{\beq}{\begin{equation}}
\newcommand{\eeq}{\end{equation}}
\newcommand{\beqn}{\begin{eqnarray}}
\newcommand{\eeqn}{\end{eqnarray}}
\newcommand{\beqno}{\begin{eqnarray*}}
\newcommand{\eeqno}{\end{eqnarray*}}
\newcommand{\bma}{\begin{displaymath}}
\newcommand{\ema}{\end{displaymath}}
\newcommand{\bnu}{\begin{enumerate}}
\newcommand{\enu}{\end{enumerate}}
\newcommand{\bce}{\begin{center}}
\newcommand{\ece}{\end{center}}
\newcommand{\btb}{\begin{tabular}}
\newcommand{\etb}{\end{tabular}}
\newcommand{\ba}{\begin{array}}
\newcommand{\ea}{\end{array}}
\usepackage{footnote}
\makesavenoteenv{tabular}
\makesavenoteenv{table}
\makeatletter 
\newcommand\semiHuge{\@setfontsize\semiHuge{21.1}{27.38}}
\makeatother

\begin{document}




\title{Flexible Unsupervised Learning for Massive MIMO Subarray Hybrid Beamforming
\thanks{This work was supported by the Natural Sciences and Engineering Research Council of Canada (NSERC) under grant RGPIN-2021-04242.}
}

\author{\IEEEauthorblockN{Hamed Hojatian,  J\'er\'emy Nadal,  Jean-Fran\c{c}ois Frigon, and Fran\c{c}ois Leduc-Primeau
}\\
{\small Department of Electrical Engineering, \'{E}cole Polytechnique de Montr\'{e}al, Montreal, Quebec, Canada, H3T 1J4}\\
\small E-mail:\{hamed.hojatian; jeremy.nadal; j-f.frigon; francois.leduc-primeau\}@polymtl.ca
}




\maketitle

\IEEEpubidadjcol

\begin{abstract}
Hybrid beamforming is a promising technology to improve the energy efficiency of massive MIMO systems. In particular,
subarray hybrid beamforming can further decrease power consumption by reducing the number of phase-shifters. However, designing the hybrid beamforming vectors is a complex task due to the discrete nature of the subarray connections and the phase-shift amounts.
Finding the optimal connections between RF chains and antennas requires solving a non-convex problem in a large search space. In addition, conventional solutions assume that perfect \gls{CSI} is available, which is not the case in practical systems. Therefore, we propose a novel unsupervised learning approach to design the hybrid beamforming for any subarray structure while supporting quantized phase-shifters and noisy CSI. One major feature of the proposed architecture is that no beamforming codebook is required, and the neural network is trained to take into account the phase-shifter quantization. Simulation results show that the proposed deep learning solutions can achieve higher sum-rates than existing methods.
\end{abstract}



\section{Introduction} \label{sec:intro}

\Gls{HBF} is a well-known approach to reduce the energy consumption of \gls{mMIMO} systems through a reduction in the number of transmitting \gls{RF} chains obtained by employing a hybrid structure combining a phase-shift \gls{AP} with a baseband \gls{DP} transmission ~\cite{hbf_survey}. As a result, \gls{HBF} techniques have been considered for fifth-generation cellular networks~(5G) in the \gls{mm-Wave} bands, and will most likely be extended to 6G~\cite{8808168}. In general, three \gls{HBF} structures are considered: \gls{FC-HBF}, \gls{FSA-HBF} and \gls{DSA-HBF}. In \gls{FC-HBF}, all RF chains are connected to all the antennas, providing the maximum degree of freedom. This structure offers the best performance, but it requires a large number of \glspl{PS} that consume a non-negligible amount of energy. In the \gls{FSA-HBF} structure, RF chains are connected to a fixed subset of antennas. The number of \glspl{PS} being reduced, the energy efficiency of this structure is higher than \gls{FC-HBF}~\cite{7436794}. Finally, in \gls{DSA-HBF}, the connections between RF chains and antennas can dynamically change using low power multiplexers, as proposed in~\cite{7880698}, and enable further energy-efficiency improvements.

However, designing \gls{HBF} that achieve near-optimal performance usually has a high computational cost. Recently, several works have investigated the use of \gls{DL} to design the \gls{HBF} for \gls{FC-HBF} \cite{alkh1, hojatian2020rssi}. To the best of our knowledge, the only \gls{DL} approach for designing subarray HBF was proposed in \cite{9189869}, for the case of a fixed subarray. However, these works proposed a supervised learning approach with perfect CSI knowledge, where the target values need to be computed using conventional \gls{HBF} algorithms that are computationally complex.
These solutions are not extendable to subarray \gls{HBF} due to additional constraints that need to be taken into account when designing the \glspl{AP}.

%

A more interesting approach consists in training a \gls{DNN} without targets, which significantly reduces the training complexity, while also making it possible to outperform the conventional approaches. An unsupervised learning method for \gls{FC-HBF} was proposed in~\cite{ unsupervised}. However, this solution is not extendable to subarray HBF because of the additional constraints that need to be taken into account when designing the \gls{AP}.
%
%
Furthermore, it is important to take into account practical system constraints to achieve the best performance. CSI must be estimated and is prone to inaccuracies, while practical phase shifters are always quantized.
Existing work address \gls{PS} quantization in two ways, either by designing with real-valued phase shifts and then applying quantization, or by constructing an \gls{AP} codebook. The codebook approach allows optimizing directly the quantized phase shifts, but the design of the codebook can be complex since the search space grows exponentially with the number of antennas, the number of RF chains, and the \gls{PS} quantization. The codebook approach also imposes a trade-off between codebook size, and the associated memory usage, and sum-rate performance.



\begin{figure*}[t!]
    \centering
    \includegraphics[width=\textwidth]{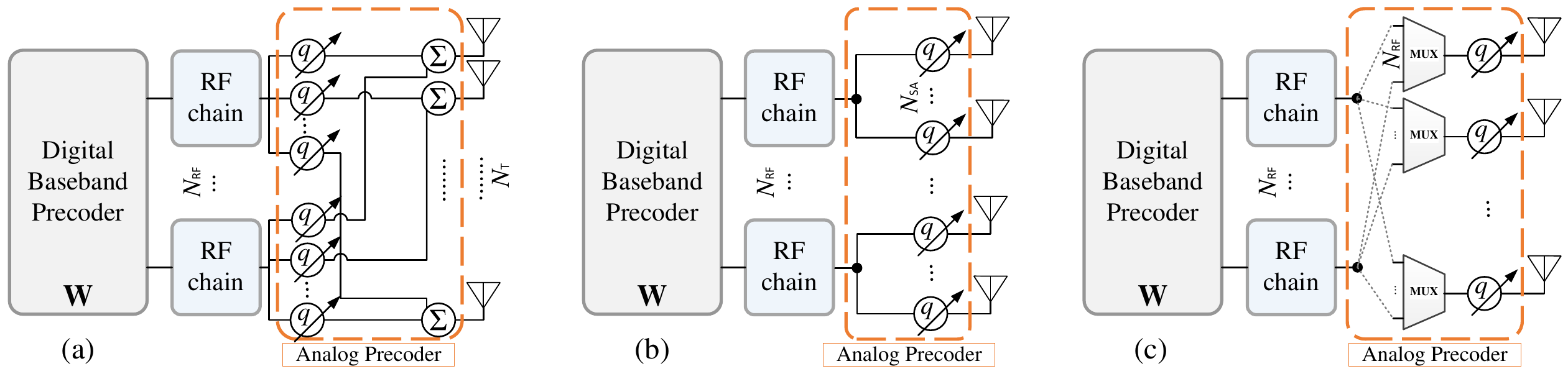}
    \caption{Hybrid Beamforming structures a) \gls{FC-HBF}, b) \gls{FSA-HBF} and c) \gls{DSA-HBF}.}
    \label{fig:HBF_arch}
\end{figure*}

 Therefore, in this paper, we propose a novel flexible unsupervised training for several \gls{HBF} structures with quantized \glspl{PS} that is not based on a codebook. Particularly, this is the first time an unsupervised DNN architecture is proposed for \gls{FSA-HBF} and \gls{DSA-HBF}. Although the DNN directly outputs quantized \gls{AP} values, the \gls{DNN} is properly trained and the gradient can be computed over the quantization layer. Since the training is unsupervised and no AP codebook is required, the amount of pre-processing steps required for DNN training has been simplified significantly. 
 Moreover, we assume that the noisy pilot signal received from the users is the only information available to the \gls{DNN} during inference. 
 We demonstrate using simulation results that the proposed solutions outperform conventional methods in all considered \gls{HBF} structures.


The rest of the paper is organized as follows. Section~\ref{Sec:sys} gives the system model. In Section~\ref{Sec:Baseline}, conventional methods are provided for each HBF structure. The proposed flexible unsupervised learning is described in Section~\ref{Sec:unsup_sol}, followed by simulation results in Section~\ref{Sec:Simulation}. Section~\ref{Sec:conclusion} concludes the paper.


\section{System Model} \label{Sec:sys}

We consider a multi-user \gls{mm-Wave} system consisting of a \gls{mMIMO} \gls{BS} in a single-cell system equipped with $N_{\sf{T}}$ antennas and $N_{\sf{RF}}$ \gls{RF} chains serving $N_{\sf{U}}$ single antenna users simultaneously. 
For both uplink and downlink transmission, \gls{HBF} precoders are employed by the \gls{BS}. We assumed that all employed \glspl{PS} have $q$ bits quantization. 

\subsection{Problem Definition}
The main objective of this paper is to design the \gls{HBF} vectors in the downlink for three different structures to maximize the sum-rate. Independently of the chosen \gls{HBF} structure, the signal received by each user can be written as
\begin{equation} \label{recv_sig}
\mb{y}_u =  \mb{h}_{u}^{\dagger} \mb{A}_{q} \sum_{\forall u}  \mb{w}_{u} x_u + \bs{\eta},
\end{equation}
where $\mb{h}_{u} \in \mathbb{C}^{N_{\sf{T}} \times 1}$ stands for the channel vector from the user index $u$ to the $N_{\sf{T}}$ antennas at the \gls{BS}, $\mb{x} = [x_1, ..., x_u, ..., x_{N_{\sf{U}}}]$ is the vectors of transmitted symbol for all users, normalized to $\mathop{\mathbb{E}}[\mb{x}\mb{x}^{\dagger}] = \frac{1}{N_{\sf{U}}} \mathbbm{1}_{N_{\sf{U}}}$, and $\bs{\eta}$ is the \gls{AWGN} term with noise power $\sigma^2$. The \gls{HBF} vectors consist of a digital baseband precoder $\mb{W} = [\mb{w}_{1}, ..., \mb{w}_{u}, ..., \mb{w}_{N_{\sf{U}}}] \in \mathbb{C}^{N_{\sf{RF}} \times N_{\sf{U}}}$ and \gls{AP} which is defined by a $N_{\sf{T}}\times N_{\sf{RF}}$ matrix $\mb{A}_{q} \in \mathbb{C}^{N_{\sf{T}} \times N_{\sf{RF}}}$, where $[\mb{A}_{q}]_{n,m}$ is the coefficient of the $q$ bits \gls{PS} connected between the $n^{\text{th}}$ antenna and $m^{\text{th}}$ RF chain. The sum-rate for a given hybrid beamformer $(\mb{A}_{q},\mb{W})$ is given by
\begin{align}
\label{eq:sumRate_HBF}
    R(\mb{A}_{q},\mb{W}) = \sum_{\forall u} \text{log}_2 \Bigl(  1+ \text{SINR}(\mb{A}_{q},\mb{w}_{u}) \Bigr) \, ,
\end{align}
where the \gls{SINR} of the $u^{\text{th}}$ user can be expressed as
\begin{align}
    \label{eq:SINR_HBF}
    \text{SINR}(\mb{A}_{q},\mb{w}_{u}) & = \frac{ \big|\mb{h}^{\dagger}_{u} \mb{A}_{q} \mb{w}_{u} \big|^2}{\sum_{j \neq u} \big|\mb{h}^{\dagger}_{u}\mb{A}_{q} \mb{w}_{j} \big|^2 + \sigma^2} \, .
\end{align}
Then, the \gls{HBF} design consists of finding the precoder matrices $\mb{W}$ and $\mb{A}_{q}$ that maximize the sum-rate in \eqref{eq:sumRate_HBF} subject to a maximum transmission power $P_{\sf{max}}$ and \gls{AP} constraints. More formally, we seek to solve the following optimization problem:
\begin{subequations}  \label{SEE-max-prb}
\begin{eqnarray} 
&\underset{\mb{A}_{q},\mb{W}}{\max} &  R(\mb{A}_{q},\mb{W}) = \underset{\mb{A}_{q},\mb{w}_u}{\max} \sum_{\forall u}  \text{log}_2 \Bigl(  1+ \text{SINR}(\mb{A}_{q},\mb{w}_{u}) \Bigr) \nonumber, \\
& \text{s.t.} & \sum_{\forall u} \mb{w}_{u}^{\dagger} \mb{A}_{q}^{\dagger}  \mb{A}_{q} \mb{w}_{u} \leq P_{\sf{max}}, \label{cnt2}
\end{eqnarray}
\end{subequations}%
where $P_{\sf{max}}$ is the normalized transmit power constraint. The analog precoder~$\mb{A}_{q}$ matrix depends on \gls{HBF} structure described in more details in the next section.

In this paper, we consider \gls{TDD} communication and we assume that the channel reciprocity is available such that the uplink channel estimate can be used for the downlink transmission.
We assume that the users transmit orthogonal pilots, such that the \gls{BS} receives
\beq
\hat{\mb{H}} = \mb{H}^{{\dagger}} + \bs{\eta},
\eeq
where $\mb{H} = [\mb{h}_{1}, ..., \mb{h}_{N_{\sf{U}}}]^{\rm{T}}$. 
As we explain in Section~\ref{Sec:unsup_sol}, the proposed \gls{DNN} directly takes as input the noisy channel estimation $\hat{\mb{H}}$.

\section{Baseline Hybrid Beamforming} \label{Sec:Baseline}
In this section we review the conventional (non-DL) solutions for each type of \gls{HBF} structure, and describe the corresponding \gls{AP}.

\subsection{Fully Connected Hybrid Beamforming~(FC-HBF)}
In \gls{FC-HBF}, as shown in Fig.~\ref{fig:HBF_arch}(a), each RF chain is connected to all $N_{\sf{T}}$ antennas through \glspl{PS} and combiners. Considering that the \glspl{PS} are quantized on $q$ bits, the set of feasible \glspl{AP} has size $2^{q N_{\sf{T}} N_{\sf{RF}}}$. Conventional \gls{HBF} solutions either consider a codebook-based solution to limit the set of feasible solutions~\cite{ayach2014} or use real-valued \glspl{PS}~\cite{yu_tsp_16}, which is not practical in a realistic system. The conventional approach consists in first designing the optimal \gls{FDP} matrix, denoted $\mb{U}_{\sf{opt}}=[\mb{u}_{1}, ..., \mb{u}_{u}, ...,  \mb{u}_{N_{\sf{U}}}]$, where $N_{\sf{RF}} = N_{\sf{T}}$. Then, the \gls{AP} and \gls{DP} are designed in such a way that the resulting precoder approximates $\mb{U}_{\sf{opt}}$ as follows:
\beq 
\underset{\mb{A}_{q}, \mb{w}_{u}}{\minimize}  \big\|\mb{U}_{\sf{opt}} - \mb{A}_{q} \mb{w}_{u}\big\|^2 \hspace{8mm}
\text{s.t. \eqref{cnt2}}, \label{eq:MMSE_prb}
\eeq
Since all the antennas are connected to all the RF chains through a \gls{PS} with $q$-bit quantization, we define $[\mb{A}_{q}]_{n,m} \in \{e^{j2\pi k/2^q} : k\in\{1,\dots,2^q\} \}$ as the feasible set for the \gls{AP}. 
The \gls{FDP} $\mb{U}_{\sf{opt}}$ in \eqref{eq:MMSE_prb} is obtained by solving the following problem:
\begin{subequations} \label{eq:FDP_opti_problem}
\begin{eqnarray} 
&\underset{\left\lbrace \mb{U}_{\sf{opt}} \right\rbrace}{\max} &
\sum_{\forall u}  R_\text{FDP}(\mb{U}_{\sf{opt}})  \\
& \text{ s. t.} & \sum_{\forall u} \mb{u}_{u}^{\dagger} \mb{u}_{u} \leq P_{\sf{max}}, \label{cnt_op2}
\end{eqnarray}
\end{subequations}
where $R_\text{FDP}(\mb{U}_{\sf{opt}}) = \sum_{\forall u} \log_2(1 + \mr{SINR}_{u}(\mb{u}_{u})) \,$ and
\begin{align}
    \label{eq:FDP_SINR}
    \text{SINR}(\mb{u}_{u}) = \frac{ \big|\mb{h}^{\dagger}_{u} \mb{u}_{u} \big|^2}{\sum_{j \neq u} \big|\mb{h}^{\dagger}_{u} \mb{u}_{j} \big|^2 + \sigma^2}.
\end{align}
The baseline results presented in this paper are obtained by solving \eqref{eq:FDP_opti_problem} based on~\cite{BBO2014},  and then obtaining the \gls{FC-HBF} solution using ``PE-AltMin'' proposed in~\cite{yu_tsp_16}. 

\subsection{Subarray Hybrid Beamforming} \label{Sec:FSA}
In subarray structures, each antenna is connected to only one RF chain through a \gls{PS}. Therefore, the total number of \glspl{PS} is reduced to $N_{\sf{T}}$ (instead of $N_{\sf{T}}\times N_{\sf{RF}}$). These structures can either have fixed connections, as shown in Fig.~\ref{fig:HBF_arch}(b), or dynamic connections (Fig.~\ref{fig:HBF_arch}(c)), where the connections between antennas and RF chains can be switched dynamically using a multiplexers network. It is shown in~\cite{7880698} that such dynamic structure improves the spectral efficiency of the system by providing more degrees of freedom in \gls{HBF} design compared to \gls{FSA-HBF}, while reducing the power consumption compared to \gls{FC-HBF}. The \gls{AP} in both subarray structures consists of two parts, i) \glspl{PS} and ii) connections between RF chains and \glspl{PS}. Therefore, the general feasible \gls{AP} matrix for subarray beamforming can be written as
\beq \label{frf-sa}
\mb{A}_{\text{q}} =\underbrace{\text{diag}\big(a_{1}, ...,  a_{N_{\sf{T}}} \big )}_{\text{Phase-shifters}} \mb{S},
\eeq
where $a_i \in \{e^{j2\pi k/2^q} : k\in\{1,\dots,2^q\} \} $, and $\mb{S}$ is a binary ${N_{\sf{T}} \times N_{\sf{RF}}}$ matrix~($[\mb{S}]_{n,m} \in \{0,1\}$) to represent the connections between the RF chains and antennas. To differentiate HBF structures, we denotes $\mb{S}_{\sf{FSA}}$ when the connection matrix $\mb{S}$ is fixed (\gls{FSA-HBF} case) and $\mb{S}_{\sf{DSA}}$ where that matrix is variable (\gls{DSA-HBF} case) and need to be jointly optimized with AP and DP. Since each antenna must be connected to only one RF chain, the constraint on matrix $\mb{S}$ is
\begin{equation}\label{DSA_cons}
      \sum_{\forall m} [\mb{S}]_{n,m} = 1 \quad \forall \ n.
\end{equation}
Some types of fixed subarray connections have been studied in~\cite{7880698}, where it was shown that type ``Squared'' achieves better sum-rate than other types. Thus, we set $\mb{S}_{\sf{FSA}}$ to the ``Squared'' connection pattern. The general approach described in \eqref{eq:MMSE_prb} can be considered here as well. In~\cite{9189869}, a solution called ``CR-AltMin'' proposed to design the \gls{FSA-HBF}. We used this method as a point of comparison for our \gls{DL} solution. 

For the \gls{DSA-HBF}, $\mb{S}_{\sf{DSA}}$ needs to be optimized, resulting in a large design space. In \cite{7880698}, a sub-optimal method named ``Dynamic Subarray Partitioning'' is proposed. 
We consider this method as a point of comparison for our proposed unsupervised learning solution.

\begin{figure*}[t!]
    \centering
    \includegraphics[width=\textwidth]{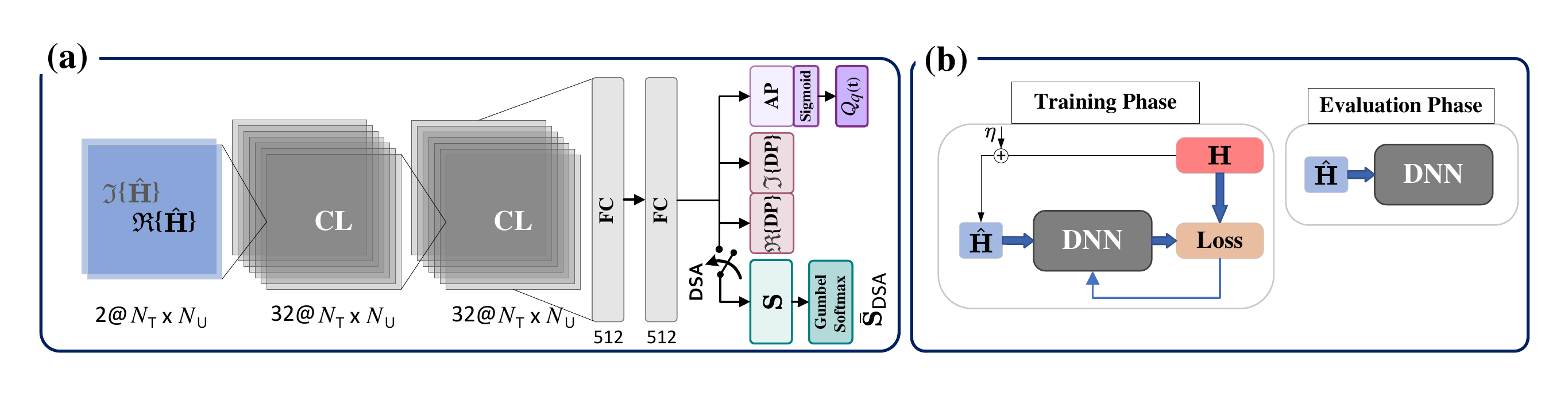}
    \caption{Proposed unsupervised learning hybrid beamforming design  a) DNN architecture, b) Training and evaluation procedure of DNN. }
    \label{fig:DNN_arch}
\end{figure*}


\section{Proposed Flexible Unsupervised Training for Subarray Hybrid Beamforming} \label{Sec:unsup_sol}
In this section, we describe a \gls{DL} architecture for designing the \gls{HBF} for fully connected and subarray structures. The training~(offline) and evaluation~(online) phases are detailed. All proposed \gls{DL} techniques are unsupervised. This is challenging because, in contrast with supervised \gls{DL}, training the DNN for each subarray structure requires different algorithms, since different hardware constraints must be considered. 


\subsection{Network Architecture}
The proposed \gls{DNN} architecture is shown in Fig.~\ref{fig:DNN_arch}~(a) and consists of $2$ \glspl{CL} $32 @ N_{\sf{T}} \times N_{\sf{U}}$ where $32$ is the number of channel and $N_{\sf{T}} \times N_{\sf{U}}$ is the dimension of each channel. The kernel size is $3 \times 3$ for all \glspl{CL} . The \glspl{CL} are followed by $2$ \glspl{FL}, each with $512$ neurons, and by an output layer. The ``Leaky ReLU'' activation function is employed after all layers, except the output layer. Batch normalization is used after each layer to avoid over-fitting. The input of the \gls{DNN} is the noisy channel matrix $\hat{\mb{H}}$ as described in Section~\ref{Sec:sys}. To improve the representation learning, we separate the real and imaginary part of $\hat{\mb{H}}$, respectively denoted $\Re\{\hat{\mb{H}}\}$ and $\Im\{\hat{\mb{H}}\}$, into two channels in the first \gls{CL}.
We divide the output of the last \gls{FL} into $4$ parallel layers. 
The output of the first parallel layer generates the AP, thus its dimension is $N_{\sf{RF}} \times N_{\sf{T}}$ for \gls{FC-HBF} and $N_{\sf{T}}$ for both \gls{FSA-HBF} and \gls{DSA-HBF}.
It is  followed by a ``Sigmoid'' activation function that scales the components of the tensor in the range $[0,1]$, and then by the quantization function~$Q_{q}$ that outputs the phase of the \glspl{PS}, defined as:
\beq \label{Q_f}
Q_{q}(t) = 2 \pi \lceil t 2^{q} \rceil / 2^{q} \quad \,
\eeq
where $\lceil x \rceil$ denotes the smallest integer greater or equal to $x$ and $t$ is an activation value.
The second and third parallel layers, both of size ${N_{\sf{RF}} \times N_{\sf{U}}}$, respectively generate the real and imaginary part of the \gls{DP}. Finally, the fourth layer of size ${N_{\sf{RF}} \times N_{\sf{T}}}$ is only activated for the \gls{DSA-HBF} structure and designs the multiplexer network. 
We considered the ``Gumbel-Softmax'' activation function~\cite{Gumbel} to satisfy the constraint described in \eqref{DSA_cons}. More details will be given in Section~\ref{sec:DSA_HBF_NET} regarding this choice. 

\subsection{Unsupervised Training}
The training phase procedure is shown in Fig.~\ref{fig:DNN_arch}~(b) (labeled ``Training Phase''). 
During training, we assume the \gls{CSI} is only available to measure the performance of the DNN through the loss function. The input of the DNN only has access to a noisy version of the \gls{CSI}.
Therefore, the training dataset is only composed of $\mb{H}$ matrix samples. 
These samples can be collected, for instance, during an initial ``offline'' phase where the channel environment is sounded. During this phase, more bandwidth resources can be dedicated toward CSI acquisition to improve the estimation accuracy.

In the training phase~(offline), the weights and biases are tuned based on a defined loss function, and back propagation computes the gradient over each layer. The use of quantization functions during training is forbidden because it is not differentiable and therefore breaks the back propagation graph. To solve this issue, \cite{9681824} proposes to use a differentiable activation function for $1$ bit \glspl{PS}. However, this method does not scale well to a higher number of bits, as we discuss later in Section~\ref{Sec:Simulation}. In this paper, we instead rely on the straight through estimator~(STE) technique~\cite{YinLZOQX19}. The quantization function~\eqref{Q_f} is only activated in the forward pass, and the output activation values are constrained in the range $[0,2\pi]$.
In the backward pass, the gradient is simply passed through unchanged.
In what follows we describe the training of each structure.
 
\subsubsection{Fully Connected Hybrid Beamforming~(FC-HBF-Net)}
In \gls{FC-HBF}, as we described in Section~\ref{Sec:Baseline}, $[\mb{A}_{q}]_{n,m} \in \{e^{j2\pi k/2^q} : k\in\{1,\dots,2^q\} \}$. Therefore, considering that $\bar{\mb{A}}_{q}$ and $\bar{\mb{W}}$ are the output of the quantized \gls{AP} and \gls{DP}, respectively, the loss function can be written as
\beq \label{loss_FCHBF}
\mathcal{L} = - R(\bar{\mb{A}}_{q},\bar{\mb{W}}),
\eeq
where $R$ denotes the sum-rate defined in~\eqref{eq:sumRate_HBF}.

\subsubsection{Fixed Subarray Hybrid Beamforming~(FSA-HBF-Net)}
The loss function used for the \gls{FSA-HBF} structure is also given by \eqref{loss_FCHBF}. However, $\bar{\mb{A}}_{q}$ is constrained to be of the form given by \eqref{frf-sa}, where $\mb{S}=\mb{S_{\sf{FSA}}}$, and $\mb{S_{\sf{FSA}}}$ corresponds to the ``Squared'' connection defined in~\cite{7880698}.

\subsubsection{Dynamic Subarray Hybrid Beamforming~(DSA-HBF-Net)}
\label{sec:DSA_HBF_NET}

The training procedure of \gls{DSA-HBF} is more challenging than the two previous structures since the matrix $\mb{S_{\sf{DSA}}}$ becomes an additional variable to be jointly optimized with the $\mb{A}_q$ and $\mb{W}$ matrices.
According to the subarray constraint defined in \eqref{DSA_cons}, each antenna must be connected to only one RF chain at any given time. Note that each RF chain can still be connected to multiple antennas.
Therefore, $\mb{S_{\sf{DSA}}}$ can be seen as a concatenation of $N_{\sf{T}}$ ``one-hot'' vectors, each of length $N_{\sf{RF}}$, where a ``one-hot'' vector is defined as a binary vector with a Hamming weight of $1$.
In classification tasks, the need to choose one of many categories is typically handled by applying the ``softmax'' function during training, which is replaced by a (hard) maximum during the test phase. However, we found that this approach does not lead to good results for unsupervised learning, because the sum rate measured during training can then be very different from the actual test-time sum rate.
To solve this issue, we propose a differentiable approximation during training inspired by the ``Gumbel-Softmax'' estimator~\cite{Gumbel}. 
Gumbel-Softmax  is a technique that allows sampling from a categorical distribution during the forward pass of a neural network, by combining the reparameterization trick and smooth relaxation. 
Defining $p_{i,j}$ the probability that antenna $i$ is connected to RF chains $j$, we can form a $1 \times N_{\sf{RF}}$ vector $\mb{p}_i = [p_{i,1},..., p_{i,N_{\sf{RF}}}]$ that corresponds to the probability states between antenna $i$ and all RF chains. Note that $\sum_j p_{i,j} = 1 \; \forall i \in [1,N_{\sf{T}}]$ due to constraint \eqref{DSA_cons}. The Gumbel-Softmax function  applied to that vector,  $G(\mb{p}_i)$, can then be defined as
\beq
 G(\mb{p}_i) = \frac{\exp((\log(\mb{p}_i) + g_i)/\tau)}{\sum_{j=1}^{N_{\sf{RF}}} \exp((\log(p_{i,j}) + g_{i,j})/\tau)},
\eeq
where $g_{i,j}$ is a sample drawn from Gumbel distribution~\cite{Gumble2} with mean $0$ and variance $1$.
Note that the $\exp(.)$ and $\log(.)$ functions are applied element-wise when taking a vector as input.
The parameter $\tau$ is called ``Softmax temperature''. When $\tau \rightarrow 0$, $G(\mb{p}_i)$ tends to the categorical distribution, but when $\tau \rightarrow \infty$, it converges to the uniform distribution~\cite{Gumbel}.
There is a trade-off between small temperatures, where sample vectors are close to one-hot but the variance of the gradient is large, and large temperatures, where samples are more uniform but the variance of the gradient is small. We thus consider $\tau$ as a hyper-parameter to be optimized. 
Finally, the output of the fourth layer, denoted $\bar{\mb{S}}_{\sf{DSA}}$, is obtained after a row-concatenation of vectors $G(\mb{p}_i) \; \forall i \in [1,N_{\sf{T}}]$. The analog precoder $\bar{\mb{A}}_{q}$ is obtained from \eqref{frf-sa}, where $\mb{S}$ is replaced by $\bar{\mb{S}}_{\sf{DSA}}$, and the loss function \eqref{loss_FCHBF} can be computed. For the power constraint, we assumed a fixed transmit power for each RF chain. Then, the power of each RF chain is split equally among the antennas it is connected to. Furthermore, the total transmit power by \gls{AP} and \gls{DP} is normalized to satisfy the power constraint in \eqref{cnt2}.

\subsection{Evaluation Phase} 

In the evaluation phase, the \gls{DNN} input consists only of noisy channel matrices~$\hat{\mb{H}}$. In all three \gls{HBF} structures, the output of the \gls{AP} is quantized based on \eqref{Q_f}, whereas the \gls{DP} can be used directly since it is continuous. For the \gls{DSA-HBF} structure, the multiplexer network in evaluation phase can be obtained by
\beq \label{one-hot}
[\mb{S}_{\sf{DSA}}]_{i,j} = 
\begin{cases}
1, & \text{if $j = \argmax_{j'} ( p_{i,j'})$,}\\
0, & \text{otherwise.}
\end{cases}
\eeq
The form of \eqref{one-hot} is known as ``one-hot encoding,'' and it determines the dynamic connection between the antennas and RF chains while ensuring that \eqref{DSA_cons} is satisfied.


\begin{table}[t!]
    \centering
    \caption{Sum-Rate Comparison of Proposed method}
    \resizebox{\columnwidth}{!}{
    \begin{tabular}{l|ccccc}
        \toprule
        \multicolumn{1}{c}{}& 
        \multicolumn{1}{c}{}&
        \multicolumn{1}{c}{Beamforming}&
        \multicolumn{1}{c}{}&
        \multicolumn{1}{c}{Sum-Rate}\\
        \multicolumn{1}{c}{Technique} & {CSI Status} & {structure} &  {\# Phase-Shifter}  & \multicolumn{1}{c}{(bit/s/Hz)} \\
        \cmidrule(lr){1-1} \cmidrule(lr){2-2} \cmidrule(lr){3-3} \cmidrule(lr){4-4} \cmidrule(lr){5-5}
        
         O-\gls{FDP}  & Perfect CSI & \gls{FDP} & - & $22.8$ \\
         \cmidrule(lr){1-1} \cmidrule(lr){2-5}
         \textbf{FC-HBF-Net}  & \textbf{Noisy CSI} & \gls{FC-HBF} &  $N_{\sf{RF}}N_{\sf{T}}$ & \textbf{21.7}  \\
         MO-AltMin~\cite{yu_tsp_16} & Perfect CSI & \gls{FC-HBF} & $N_{\sf{RF}}N_{\sf{T}}$ & $18.9$ \\
         PE-AltMin~\cite{yu_tsp_16}  & Perfect CSI & \gls{FC-HBF} & $N_{\sf{RF}}N_{\sf{T}}$ & $17.8$ \\
         OMP~\cite{ayach2014} & Perfect CSI & \gls{FC-HBF} &  $N_{\sf{RF}}N_{\sf{T}}$ & $15$ \\
         \cmidrule(lr){1-1} \cmidrule(lr){2-5}
         \textbf{DSA-HBF-Net}  & \textbf{Noisy CSI} & \gls{DSA-HBF} & $N_{\sf{T}}$ & \textbf{20.1} \\
         Park, et al~\cite{7880698}  &  Perfect CSI &\gls{DSA-HBF} & $N_{\sf{T}}$ & $18.7$ \\
         \cmidrule(lr){1-1} \cmidrule(lr){2-5}
         \textbf{FSA-HBF-Net} & \textbf{Noisy CSI} & \gls{FSA-HBF} &  $N_{\sf{T}}$ & \textbf{16.4}  \\
         CR-AltMin~\cite{9189869} & Perfect CSI & \gls{FSA-HBF} &  $N_{\sf{T}}$ & $13.8$ \\

        \bottomrule
        \multicolumn{3}{c}{($N_{\sf{T}} = 64$, $N_{\sf{RF}} = 8$, $N_{\sf{U}} = 4$, $\sigma^2 = -130$ dBW)}
    \end{tabular}}
    \label{tbl:comparison}
\end{table}

\section{Numerical Results} \label{Sec:Simulation}

The performance of the proposed solutions have been evaluated numerically by using the \textsc{PyTorch} \gls{DL} framework. Scenario ``O1-$28$~GHz'' of the deepMIMO channel model~\cite{deepmimo} is employed to generate the dataset. The \gls{BS} is equipped with $N_{\sf{T}} = 64$ antennas and $N_{\sf{RF}} = 8$ RF chains with $q$-bit  \glspl{PS} serving $N_{\sf{U}}=4$ users located randomly in a dedicated area~(in deepMIMO channel model $\texttt{active\_user\_first} = 1100$ and $\texttt{active\_user\_last} = 2200$). 
The size of the DNN dataset is set to $2 \times 10^{6}$ samples, with $85$\% of the samples used for the training set and the remaining ones used to evaluate the performance. The mini-batch size, learning rate, and weight decay are set to $1000$, $0.001$, and $10^{-5}$, respectively. We used ``RAdam'' as the DNN training optimizer~\cite{LiuJHCLG020}. The training procedure converged after approximately $200$ epochs.

Table~\ref{tbl:comparison} compares the sum-rate performance of all techniques. The best sum-rate is obtained by O-FDP since the RF structure is not constrained, but this solution is not energy-efficient compared to HBF. 
The DNN architecture implementing the \gls{FC-HBF} structure achieves $95 \%$ of \gls{FDP} sum-rate, while improving the sum-rate by $15\%$ when compared to the MO-AltMin solution~\cite{yu_tsp_16}.
FSA-HBF-Net outperforms existing methods by $19\%$. Furthermore, the DSA-HBF-Net approximately reaches $92\%$ of \gls{FC-HBF} sum-rate, while the number of \glspl{PS} has been divided by a factor $N_{\sf{RF}}$. For DSA-HBF-Net, the Gumbel-Softmax temperature is set to $\tau = 1.5$. This choice will be motivated later in this section. When compared to  ``Dynamic Subarray Partitioning'' in \cite{7880698}, the proposed approach improves the sum-rate by $7\%$.
Note that all conventional methods were evaluated using perfect CSI, whereas the proposed DNN techniques consider noisy CSI in the evaluation phase. Therefore, it is expected that the performance gap would be larger in practice.

\begin{figure}[t!]
    \centering
    \includegraphics[width=\columnwidth]{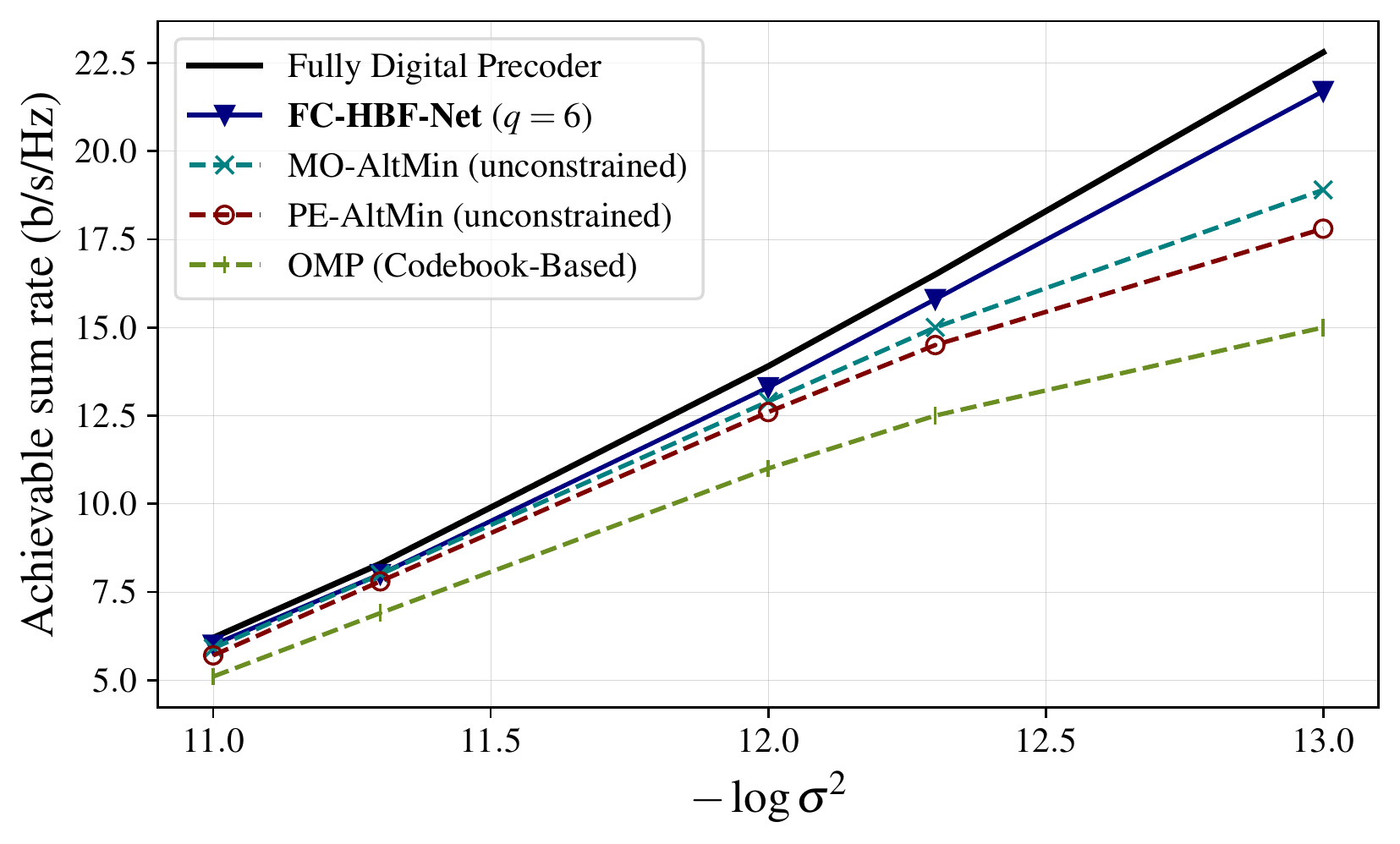}
    \caption{Sum-rate performance of \gls{FC-HBF}.}
    \label{fig:SR_FC}
\end{figure}

Fig.~\ref{fig:SR_FC} shows the sum-rate performance of FC-HBF based solutions for different noise power values $\sigma^2$. Sum-rate of the optimal FDP solution is also shown and provides an upper-bound for the sum-rate performance. Taking into account channel attenuation, the average \glspl{SNR} ranges from $6.7$ dB to $26.7$ dB. The proposed FC-HBF-Net outperforms all conventional FC-HBF solutions for all noise power. Note that the performance gap between \gls{FDP} and \gls{HBF} increases at low noise power due to residual interference being more dominant for HBF systems.

\begin{figure}[t!]
    \centering
    \includegraphics[width=\columnwidth]{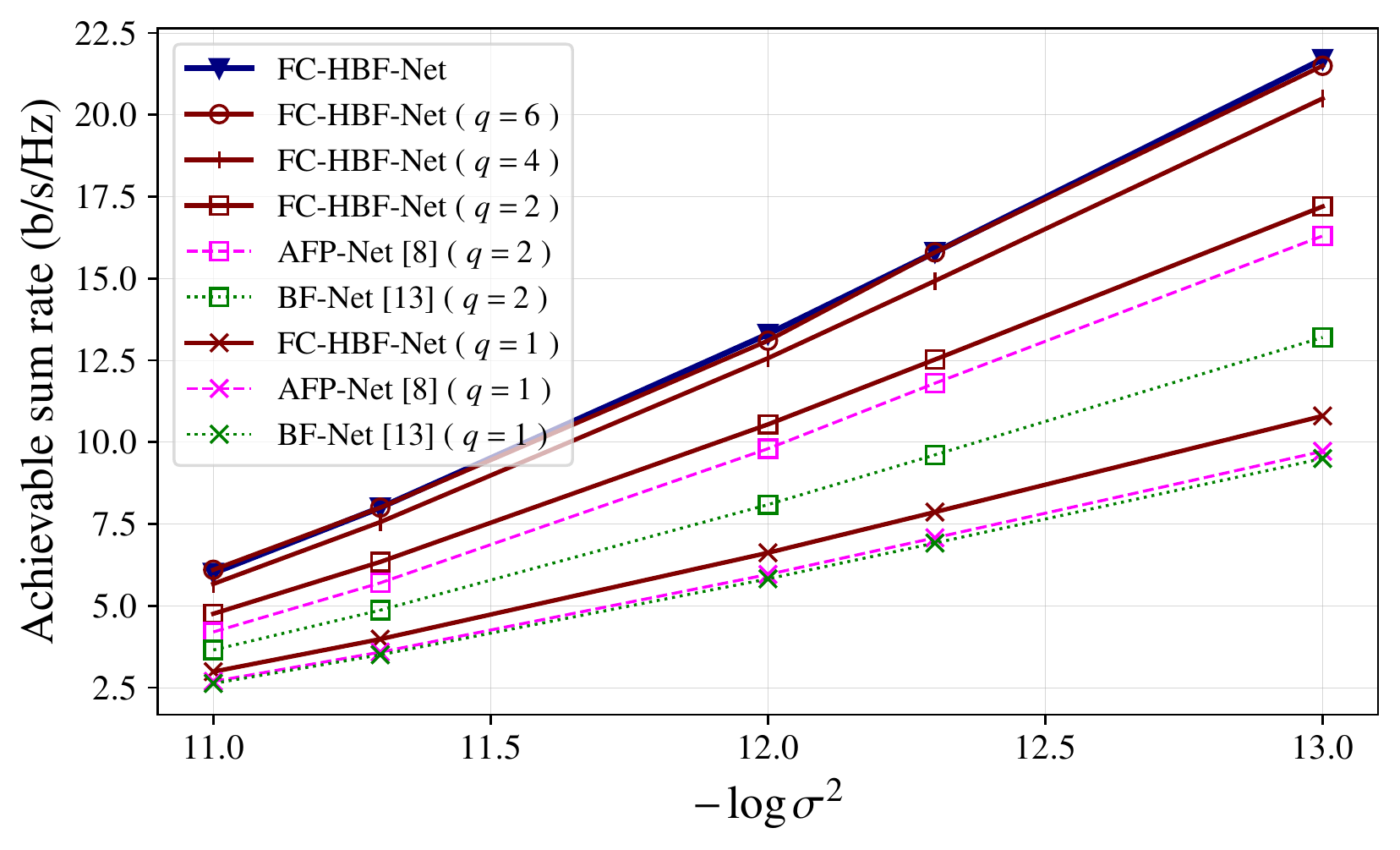}
    \caption{Sum-rate of DL-Based HBF with different \gls{PS} quantization bits.}
    \label{fig:SR_bits}
\end{figure}

Fig.~\ref{fig:SR_bits} shows the impact of the \gls{PS} quantization on the sum-rate performance.  
We consider PS representations from 1 to 6 bits. At 6 bits, the sum-rate becomes almost identical to the ideal case of continuous phase shifts. 
As a comparison for the proposed approach, Fig.~\ref{fig:SR_bits} also reports the sum-rate achieved by the codebook-based solution ``AFP-Net''~\cite{unsupervised}, for 1 and 2-bit PS quantization. It can be seen that the codebook-based solution provides lower sum-rate since its performance is limited by the codebook design.
In contrast, the proposed DL solution directly finds the optimal solution.
We also compare against an existing 1-bit quantization method trained using a differentiable function~\cite{9681824}, that
we adapted to also support a 2-bit quantization.
These reference curves are labeled ``BF-Net'' in Fig.~\ref{fig:SR_bits}.
To evaluate the performance on our dataset, we replaced our quantization function with the one from \cite{9681824}, taking care to re-optimize the method's $\alpha$ parameter since it could depend on the DNN architecture and channel model. 
We observed that $\alpha=0.1$ is the best trade-off between the training and evaluation loss for $q \in \{1,2\}$.
It can be seen that the use of the STE quantization trick allows the proposed \gls{DNN} to be trained properly while supporting any \gls{PS} quantization, and that our approach outperforms the quantization method in \cite{9681824}. Particularly, the performance gap is significantly higher at $q = 2$, which tends to show that the method proposed in \cite{9681824} does not scale well when increasing the number of quantization bits.

\begin{figure}[t!]
    \centering
    \includegraphics[width=\columnwidth]{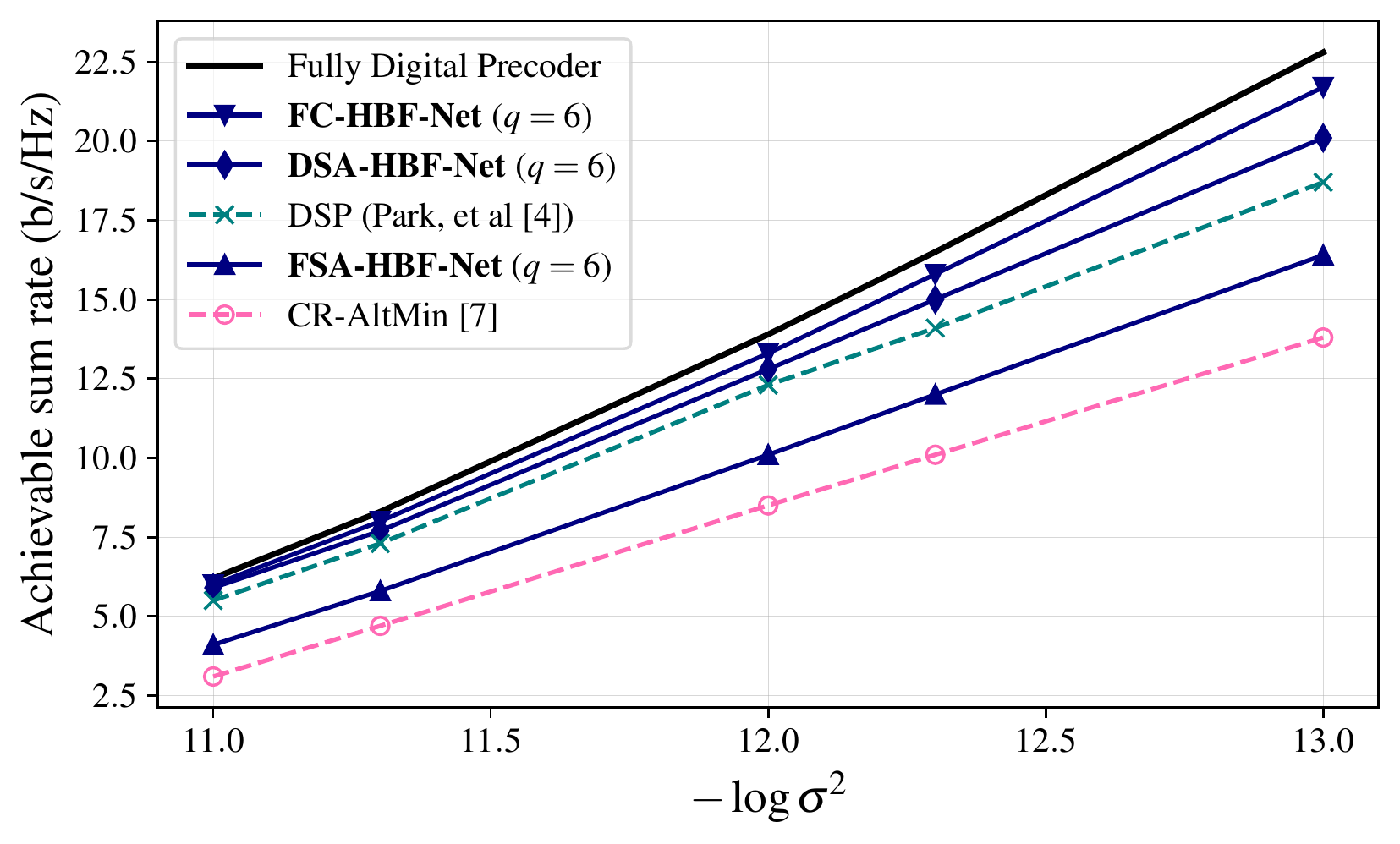}
    \caption{Sum-rate performance of \gls{FC-HBF}, \gls{FSA-HBF}, and \gls{DSA-HBF}.}
    \label{fig:SR_all}
\end{figure}

Fig.~\ref{fig:SR_all} shows the obtained sum-rate performance when considering different \gls{HBF}  structures. \gls{FC-HBF} being the less constrained structure, it achieves the highest sum-rate performance. The most constrained structure, \gls{FSA-HBF}, reduces the power consumption of the RF phase-shifter array at the cost of offering the worst sum-rate performance. The sum-rate performance of the \gls{DSA-HBF} is very close to \gls{FC-HBF} while only requiring $N_{\sf{T}}$  \glspl{PS}. The multiplexer network improves the \gls{HBF} design flexibility, which in turn improves the sum-rate when compared to \gls{FSA-HBF}. 

\begin{figure}[t!]
    \centering
    \includegraphics[width=\columnwidth]{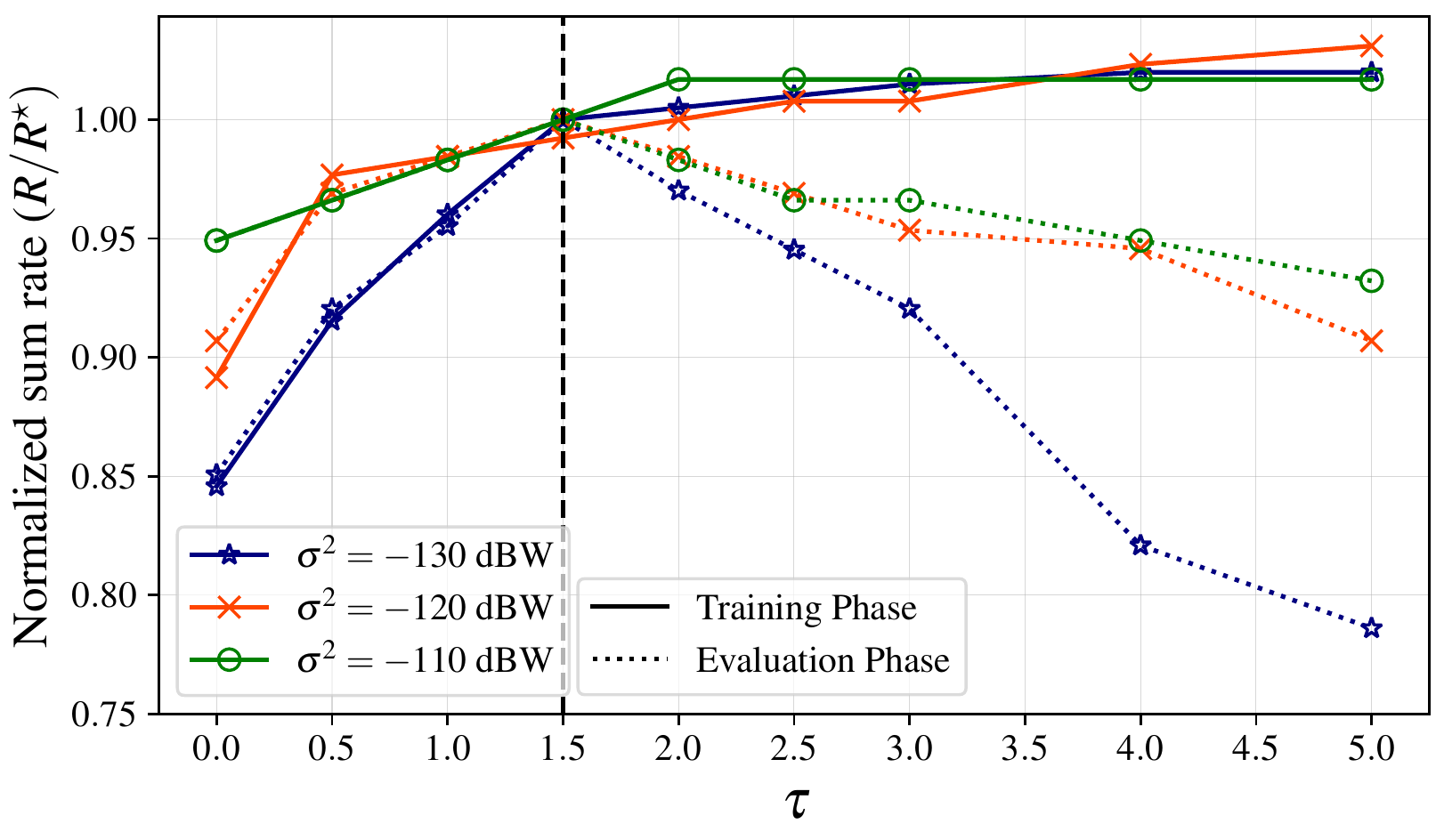}
    \caption{Sum-rate performance of \gls{DSA-HBF} versus different $\tau$.}
    \label{fig:SR_tau}
\end{figure}

Finally, we evaluate in Fig.~\ref{fig:SR_tau} how the Gumbel-Softmax temperature $\tau$ impacts the sum-rate performance of \gls{DSA-HBF}-Net during the evaluation and training phases, for different noise powers. To improve the clarity of the figure, we normalize the sum-rate using $R/R^{\star}$, where $R^{\star}$ is the maximum sum-rate achieved at each noise power during the evaluation phase. We can observe that the best evaluation-phase sum-rate among the search options is obtained at $\tau = 1.5$ for all noise powers. Note that for higher $\tau$ values, the sum-rate performance of the training phase keeps improving, while the sum-rate is significantly degraded in the evaluation phase. 
This is due to the fact that Gumbel-Softmax converges toward a uniform distribution for large $\tau$, and therefore becomes a bad representation of the desired one-hot vector.



\section{Conclusion} \label{Sec:conclusion}

 Subarray hybrid beamforming enables to further improve the energy efficiency of conventional \gls{HBF} by reducing the number of phase-shifters. We propose for the first time an unsupervised DNN architecture for \gls{FSA-HBF} and \gls{DSA-HBF}, while supporting quantized phase shifters of any resolution. Since quantization functions are not differentiable and, consequently, cannot be used in back propagation, we considered the STE technique to train the network with quantized \glspl{PS}.
 For \gls{DSA-HBF}, we proposed to implement the Gumbel-Softmax activation function to efficiently train the network while satisfying the connection constraints between RF chains and antennas. Simulation results show that the proposed unsupervised \gls{DL} techniques outperform the conventional HBF techniques for all \gls{HBF} structures, even though the system uses noisy instead of perfect CSI as input.
 


\bibliographystyle{IEEEtran}
%
\bibliography{bib/HBF_DL.bib}

\end{document}